# Polariton resonances for ultra-strong coupling cavity optomechanics in GaAs/AlAs multiple quantum wells

B. Jusserand[1], A. Poddubny[2], A. Poshakinskiy[2], A. Fainstein[3], and A. Lemaitre[4]

1. Institut des Nanosciences de Paris, CNRS UMR 7588, Université Pierre et Marie Curie (UPMC), 75005 Paris, France
2. Ioffe Institute, 194021 St. Petersburg, Russia
3. Centro Atomico Bariloche and Instituto Balseiro, C.N.E.A., 8400 S. C. de Bariloche, R. N., Argentina
4. Laboratoire de Photonique et de Nanostructures, C.N.R.S., 91460 Marcoussis, France

*Polariton-mediated light-sound interaction is investigated through resonant Brillouin experiments in GaAs/AlAs multiple-quantum wells. Photoelastic coupling enhancement at exciton-polariton resonance reaches $10^5$ at 30 K as compared to a typical bulk solid room temperature transparency value. When applied to GaAs based cavity optomechanical nanodevices, this result opens the path to huge displacement sensitivities and to novel ultrastrong-coupling cavity phenomena with optomechanical couplings $g_0$ in the range of 100 THz.*

The field of cavity optomechanics [1] offers a rich variety of novel phenomena and applications, mostly based up to now on the silicon platform in which nanofabrication techniques have reached a very high level of maturity. This technological choice has lead to consider essentially a single dominant mechanism for the coupling between optical and mechanical degrees of freedom in micro- or nanoresonators, the so-called radiation pressure. In that case, optical resonances are modified by the surface and interface displacements exclusively. More recently GaAs has been considered [2, 3] as an alternative choice of great potential in relation with the well established optoelectronic properties of direct gap semiconductors not available in silicon. It has been pointed out in Ref. [3] that, in GaAs, radiation pressure has to be combined with a second mechanism, the photoelastic coupling, in order to quantitatively describe optomechanical coupling efficiency in nanomechanical devices based on the GaAs platform. The photoelastic coupling describes the modification of the dielectric properties of the device in the presence of strain fields accompanying the mechanical behavior [4-6].

Optimizing optomechanical coupling in GaAs nanocavities thus does not rely only on increasing the confinement of optical and acoustic fields at the same location in the device, in other words in nanofabricating resonators with efficient acoustic-optic overlap and high-quality factors for both photons and phonons. In fact, it can also benefit from an optimization of the photoelastic coefficients in the constituting materials. Contrary to radiation pressure, photoelastic coupling indeed strongly depends on the wavelength of light involved in the experiments and, in particular, on its detuning from intrinsic optical resonances in the material [7]. These resonances can be described either in the excitonic or in the polaritonic picture depending upon experimental conditions such as the temperature, the residual non radiative damping or the inhomogeneous broadening of the relevant transitions [8]. As recently proposed, polaritons in cavities could allow the access of a fully resonant light-sound interaction regime, avoiding detrimental effects related to dissipation, which is of great interest for cavity optomechanics [9]. A model for the phenomena, and a determination of the magnitude of this interaction, is however still lacking.

Photoelastic coupling in bulk GaAs has been measured in the last decades using standard piezo-optics schemes such as transmission birefringence [10] or ellipsometry [11, 12] in the presence of externally applied static stress unto the sample. These experiments provide useful information either in the transparency region or in the presence of significant absorption, respectively, i.e. they leave the strong resonance domain near the absorption edge fully uncovered. Moreover, temperature dependences have usually not been considered in these experiments, as non-resonant properties only weakly depend on this external parameter. Piezo-transmission experiments under externally launched acoustic waves have been also reported for GaAs with similar results and limitations [13, 14]. Finally, resonant Brillouin scattering has been demonstrated as a powerful method to study acousto-optical coupling near exciton resonances [15, 16]. It has the additional advantage that thermally activated internal strains are involved and usually complex stress apparatus can be fully avoided. Unfortunately, a quantitative description of the resonant Brillouin scattering cross section including the exciton-polariton range had not been obtained until very recently [17].

Systematic studies of the photoelastic coupling thus emerge as of great interest in the new developing field of GaAs based cavity optomechanical nanodevices. In this Letter, we show that the analysis of the Brillouin scattering intensity in GaAs/AlAs multiple quantum wells (MQW) provides a quantitative determination in a large range of temperatures of resonant optical and optomechanical parameters with unprecedent accuracy and completeness as compared to previous piezo-optical experiments [10-14]. We demonstrate huge enhancement of the photoelastic constants of GaAs/AlAs MQWs at resonance and with decreasing temperature, thus opening the way to novel ultrastrong coupling regimes in cavity optomechanics.

We performed resonant Brillouin backscattering experiments between 30 K and 290 K on a very high quality MQW containing 40 GaAs wells with a thickness of 17.1 nm separated by AlAs barriers of 7.5 nm thickness [17]. With such relatively thick barriers, the electronic states in each quantum well have a negligible overlap with their analogues in neighboring wells. Only radiative coupling between quantum well excitons with non-dispersive character along the stacking direction governs their interaction. This system thus provides a unique realization of the model excitonic polariton theory as introduced by Hopfield [18] and by Pekar [19] more than fifty years ago. A MQW is also a periodically modulated acoustic structure with a large number of periods so that its acoustic properties can be described in the folding scheme [4]. Acoustic waves with wavevectors shifted by a multiple of the Brillouin minizone extension $2\pi/d$, and energies much larger than the standard Brillouin energy become active for polariton scattering. We have covered the largest possible energy range including both the low energy tail of the resonance towards the transparency region and the light-hole exciton range, a few meV above the heavy-hole one. When the temperature is increased, the overall decrease of the scattered intensity, as compared to other background secondary emissions, reduces the accessible energy range. We focus in this Letter on the quantitative analysis of the temperature dependent heavy-hole fundamental exciton polariton resonance which is the strongest and the best defined one along the whole accessible temperature range.

We show in the left panel of Fig. 1 typical Brillouin spectra obtained when the laser energy is very close to the heavy-hole exciton energy (strong resonance) at a few different temperatures using a very narrow and stable tunable laser line and a double spectrometer with a very high resolution close to 0.11 cm$^{-1}$ (14 µeV). When the temperature is decreased, the scattering intensity increases dramatically (not visible in Fig.1 showing rescaled spectra). It always remains sufficiently intense as compared to both competing excitonic luminescence and resonant Rayleigh line to allow accurate determinations of the shift, the width and the intensity of several Brillouin lines associated to three folded acoustic lines at least, both on the Stokes and the anti-Stokes sides of the Rayleigh line. Below 30 K, exciton luminescence strongly increases and the Brillouin lines broaden in such a way that a clear separation of the different signals becomes difficult in the most interesting strong resonance energy range. This looks counterintuitive as the non radiative exciton damping decreases in the same conditions (see the insert in Fig. 2). We will show below that at resonance, the peak Brillouin linewidth indeed increases with decreasing exciton damping while the energy range over which the width is modified strongly shrinks.

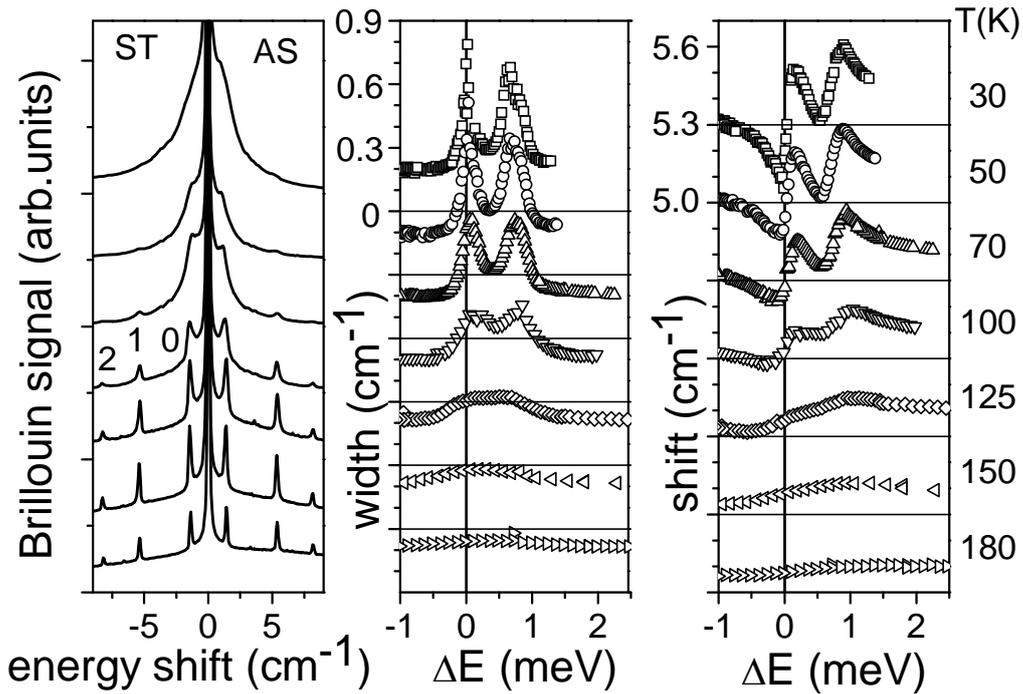

*Figure 1: Left panel: Brillouin backscattering spectra obtained at different temperatures between 30 and 180 K (indicated on the right side). Intensities have been rescaled and the baselines shifted for clarity. For each temperature, the laser energy coincides with the maximum of (heavy-hole) exciton line and comparable intensities are obtained on the Stokes and the anti-Stokes sides of the spectra. The lines are labeled 0, 1 and 2 corresponding to the unfolded LA line and the folded FLA1 and FLA2 lines respectively. Middle and right panel: Variation of the width and the shift of the FLA1 line in the Stokes side measured at the same temperatures as in the left panel, plotted as a function of the energy shift relative to the temperature dependent exciton resonance energy. The vertical line shows the exciton energy aligned with the lowest energy feature visible in both the width and the shift patterns. The baselines have been shifted by steps of 0.3 $cm^{-1}$, i.e. the separation between two successive major ticks, for clarity. In the center (right) panel, the baseline (resp. a line at 5.3 $cm^{-1}$) is shown as a thin line for each temperature.*

We show in the middle and right panels of Fig.1 the variation with incident energy of the shift and the width of the FLA1 line, for temperatures ranging between 30K and 180K. The measured curves have been shifted vertically to show relative variations around the temperature dependent heavy-hole exciton energy. Two structures are visible in the energy dependence of both the shift and the width of the Brillouin peaks. They reflect the exciton polariton dispersion when either the incident laser energy or the scattered energy become close to the exciton resonance [8]. The temperature dependence demonstrates the formation of the polariton gap when the temperature is decreased, i.e. when the non radiative damping is decreased. At 30K, the resonant dependence of the Brillouin linewidth becomes very narrow, with maximum values around 0.8 cm$^{-1}$, while the Brillouin shift oscillations become very abrupt. The polaritonic behavior remains visible until 125-150K while at 180K, a standard non-dispersive behavior with a constant shift and a very small width is recovered. Nevertheless, the intensity variations give sufficient information to fit the data up to room temperature. Our model provides a comprehensive description of the results. It also gives access to the relevant cavity optomechanics parameters, namely the resonant photoelastic coupling constants and photon lifetimes.

We show in Figure 2 the intensity variations of the FLA1 Stokes line for all temperatures available in this study. We used the lowest possible laser intensities to measure the resonance behaviors. Values ranged between 10µW at low temperature and strong resonance and 2mW at high temperature. The intensities measured for different temperatures have been carefully normalized to each other and only a global intensity factor common to all data remains arbitrary. As a calibration of the spectrometer response we have also measured the intensity variation of the LO line of silicon and obtained a maximum variation of 25% over the full range of energies relevant to this study, a negligible variation with respect to the several orders of magnitude observed for the resonant multiple quantum well Brillouin lines studied in this work.

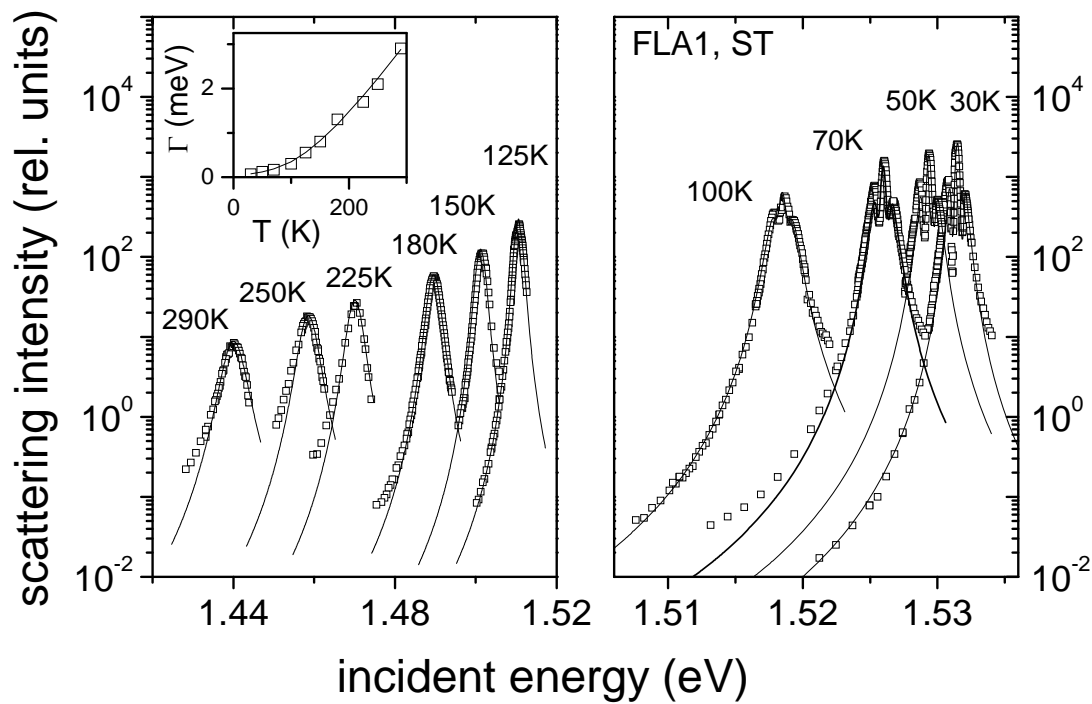

*Figure 2 Variation of the FLA1 line intensity in the Stokes side measured at different temperatures between 30 and 290K (open squares), plotted as a function of the incident laser wavelength. The latter has been varyied around the temperature-dependent exciton resonance. The horizontal scale is different in the two panels. The full lines represent the best fit of the experimental data. The temperature variation of the non-radiative damping parameter (open squares) deduced from the fit is shown in the inset and compared with a simple formula taking into account both the contributions of acoustic and LO phonon to the damping (full line).*

In the same figure, we also show the best fit obtained using the polariton model described in Ref. [17]. In this fit, all the parameters have been taken constant as a function of the temperature except for the exciton energy and its non-radiative damping. The radiative damping has been taken as 37 µeV and the non radiative damping evolution is shown in the inset. It increases regularly with temperature and its value can be remarkably well fitted with a standard formula taking into account a linear contribution associated to the acoustic phonon thermal population and a second contribution proportional to the LO phonon Bose Einstein population ($n_{LO}$), $\Gamma(T)$ [meV] = 2.4 x $10^{-3}$ [meV/K] T + 7.0 [meV] $n_{LO}$. Both fitted coefficients are in excellent agreement with previous determinations [19, 20]. Quite notably, a negligible constant offset is deduced from the fit, that is $\Gamma(0) = 0$. In contrast, non vanishing temperature independent contributions have been always deduced from previous experiments, mostly performed on single quantum wells. As we would have assumed a larger contribution in multiple quantum wells due to well fluctuations, our result illustrates both the excellent sample quality and the MQW polariton collective nature. This latter mechanism averages out, in a similar behavior to motional narrowing [21], the roughness and small layer thickness fluctuations always present in the different quantum wells included in the sample.

As demonstrated by Fig.2, based on the theory developed in [17] we obtain a comprehensive quantitative description of the measured intensities in a GaAs/AlAs MQW from room temperature, when the non-radiative damping dominates and washes out the polariton features, down to 30K, where very clear dips appear in the scattering intensity curves when the incident or the scattered energies are tuned to the polariton gap. Notably, at 30K the non radiative contribution to the linewidth becomes as low as 70 µeV, only twice the value obtained for the radiative one. Brillouin scattering has been previously used to determine photoelastic coefficients in semiconductors near excitonic resonances but the measurements remained limited to the transparency side of the resonance [22, 23]. Based on the successful quantitative modeling of our measurements over the whole energy range, we extended this concept to our multi-quantum-well and to the polaritonic description developed in this work. The calculation procedure is outlined below.

The quantum well photoelastic response can be characterized by the fourth-rank frequency-dependent tensor $p_{ijkl}(\omega_l, \omega_s)$ linking the Fourier components of the dielectric polarization $P_i(\omega_s)$, the electric field $E_j(\omega_l)$ and the strain tensor $u_{kl}(\omega_s - \omega_l)$:

$$4\pi P_i(\omega_s) = -\varepsilon_b^2 \int \frac{d\omega_l}{2\pi} p_{ijkl}(\omega_l, \omega_s) E_j(\omega_l) u_{kl}(\omega_s - \omega_l) \tag{1}$$

Here, $\varepsilon_b$ is the quantum well background dielectric constant and *i*, *j*, *k* and *l* are Cartesian indices. The difference between of the incident wave frequency $\omega_l$ and that of the scattered wave $\omega_s$ is due to the absorption (emission) of the phonon with the frequency $\Omega = |\omega_s - \omega_l|$. We study the case when the incident and scattered waves propagate normally to the quantum well surface, lying in the *xy* plane, and the acoustic phonons are of longitudinal type. In this case *i=j=x*, and *k=l=z*. The relevant component of the photoelastic tensor [$p_{12}(\omega) = -\frac{1}{\varepsilon_b^2} \frac{\partial \varepsilon_{ii}(\omega)}{\partial u_{jj}}$ with $i \neq j$] describes the modification of the coherent dielectric response under the deformation. It can be directly extracted from the generalized photoelastic function $p_{ijkl}(\omega_l, \omega_s)$ defined in Eq. 1, as $p_{12}(\omega) = p_{12}(\omega, \omega)$. In fact, Brillouin scattering provides the tool to directly probe the photoelastic response $p_{12}(\omega)$ as an extrapolation to Ω=0 of measurements of $p_{12}(\omega_l, \omega_l \pm \Omega)$ at a few different values of Ω both in Stokes and anti-Stokes components of the light scattering spectra. In the vicinity of an excitonic resonance in a single quantum well at the frequency $\omega_0$ the function $p_{12}(\omega_l, \omega_s)$ assumes the form [7]:

$$p_{12}(\omega_l, \omega_s) = \frac{4\pi \xi \Gamma_0 \Xi(\Omega)}{\varepsilon_b^2 (\omega_0 - \omega_l - i\Gamma)(\omega_0 - \omega_s - i\Gamma)}, \tag{2}$$

where $\xi = c\sqrt{\varepsilon_b}/(2\pi a \omega_0)$ with *a* being the quantum well width, $\Gamma_0$ is the exciton radiative decay rate, $\Gamma$ the nonradiative damping and $\Xi(\Omega)$ the matrix element of the deformation potential weighted by the overlap integral between the light wave and the phonon mode, corresponding to the frequency $\Omega$. Equation (2) is applicable provided that the nonradiative damping of the exciton significantly exceeds the radiative one ($\Gamma > \Gamma_0$). This is the case even for the lowest considered temperature, *T*=30 K (where $\Gamma = 2\Gamma_0$).

The scattering spectrum $I(\omega_l, \omega_s)$ can be written as a sum over the contributions from different quantum wells [17]. The result is expressed in terms of the generalized photoelastic function as follows:

$$I(\omega_l, \omega_s) \propto T |t(\omega_l) t(\omega_s)|^2 |\Delta[Q(\omega_l) + Q(\omega_s) - k]|^2 |p_{12}(\omega_l, \omega_s)|^2. \quad (3)$$

The function $\Delta(Q) = \sum_{n=1}^{N} e^{iQz_n}$ is the structure factor, where the summation runs over the $N$ quantum wells located at the points $z_n$. $k = \pm \Omega/s$ is the phonon wavevector, where the sign +(-) corresponds to LA, FLA2 (FLA1) modes respectively. $Q(\omega) = \omega/c \sqrt{\varepsilon_{eff}(\omega)}$ is the wave vector of the excitonic polariton, $t(\omega) = 2q/[q + Q(\omega)]$ is the transmission coefficient with $q = \omega/c \sqrt{\varepsilon_b}$, and $T$ is the temperature. For this purpose, and as the considered superlattice period is significantly smaller than the light wavelength, the polariton dispersion can be well approximated by the effective dielectric constant:

$$\varepsilon_{eff}(\omega) = \varepsilon_b \left( 1 + \frac{\omega_{LT}}{\omega_0 - \omega - i\Gamma} \right), \quad (4)$$

in which $\omega_{LT}$ is the longitudinal transverse splitting of the polariton.

Equation 3 allows us to derive the model parameters based on the simultaneous fit of six different resonant Brillouin intensity variations at each temperature. We show in Figure 3 two different examples of this fit at the extreme studied temperatures of 30K (left panel) and 290K (center panel). At 290K, the fit is excellent. The resonances curves have no structures due to the large non radiative damping of the exciton at this temperature, which washes out the polariton effects. Nevertheless, the position of the maximum displays a small but significant variation from line to line. This variation observed for different $\Omega$ shows that the so-called static approximation in which the difference between $\omega_l$ and $\omega_s$ is neglected does not apply even at room temperature. At 30K, the fit remains very good for the folded lines but shows imperfections for the LA line at energies close to the exciton. This could be due to some experimental uncertainties as the exciton luminescence and the resonant Rayleigh line at the laser energy become very intense in these conditions and can perturb the lineshape fitting of the LA lines.

We also show in Figure 3 the calculated static ($\Omega$=0) intensities based on the same parameters used to fit the Brillouin line intensities. They display resonant profiles very similar to the ones for acoustic lines, with somewhat larger peak values due to the superposition of the Stokes and anti-Stokes contributions for $\Omega$=0. We show in the top right panel of the figure the static function $p_{12}(\omega)$ deduced with the same approach at all available temperatures. Brillouin scattering provides the modulus of this quantity which is real below the absorption edge but is a complex quantity above it, as it is a strain derivative of the dielectric function, which is a complex quantity above the absorption edge [24]. In order to calibrate our determination of $p_{12}(\omega)$, we compare our data with an independent determination of $p_{12}(\omega)$ in the transparency region [13, 14]. The latter has been done on bulk GaAs and not on a MQW. The exciton binding energies and oscillator strengths are slightly smaller in the bulk. Moreover, the light hole and the heavy hole are not split as they are by confinement in QW. As we observed in our sample that the contribution of the light hole and the electron-hole continuum to the resonant intensities are quite small as compared to the one of heavy hole, we conclude that our comparison of the QW heavy hole contribution with the bulk exciton one should give a reasonable estimation of the absolute photoelastic function in our sample. As we handle with several orders of magnitude variations as a function of temperature and energy distance to the exciton, this estimation should be accurate enough for future use in the analysis of GaAs MQW optomechanical experiments.

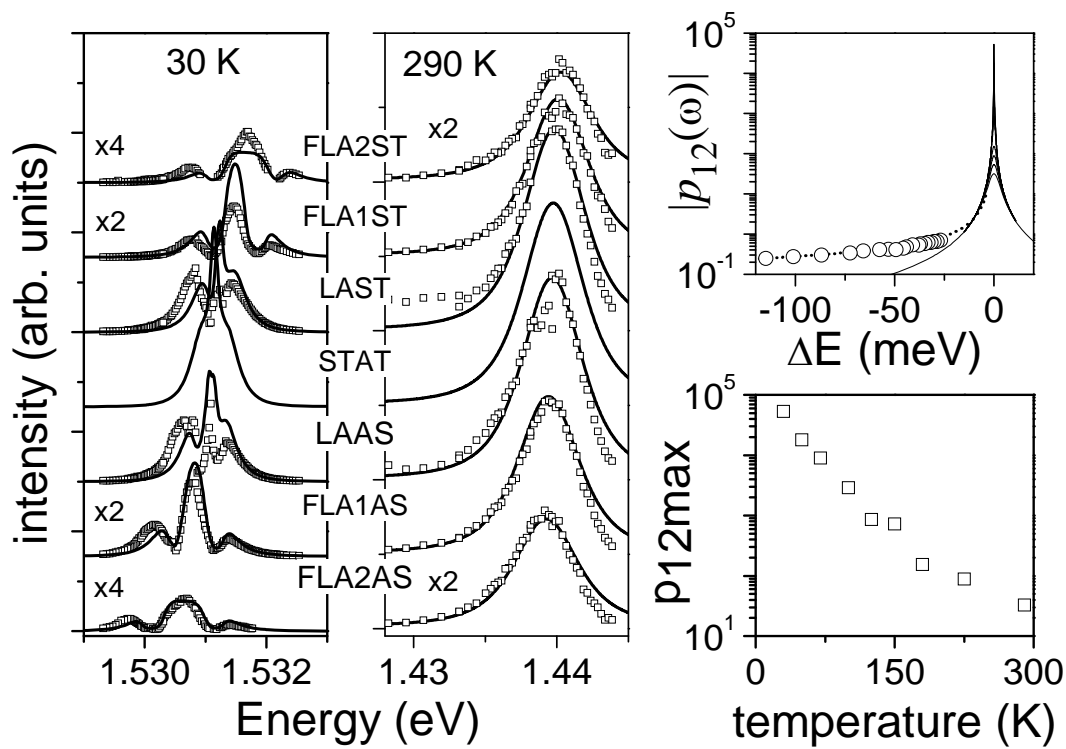

Figure 3 Variation of the intensity of all Brillouin line measured at 30K (left panel) and 290K (center panel), plotted (open squares) as a function of laser energy, around the fundamental

*exciton resonance (different energy scale for the two temperatures). The continuous curves represent the best fit of the experimental data. The calculated static component (Ω=0) is also shown in both panels. The top right panel shows in thin lines the modulus of $p_{12}(\omega)$ deduced from the static component in our model for all considered temperatures. The curve corresponding to 290K is used to provide an absolute calibration for $p_{12}(\omega)$ by comparison with a previous determination performed at this temperature in the transparency region [13, 14], shown with open circles. To allow for an energy overlap between the two measurements, a theoretical extrapolation of these data towards the exciton energy is also presented (dotted line). Based on this absolute calibration the resulting maximum values of $p_{12}$ at resonance are shown in the bottom right panel for the different temperatures measured (see text for more details).*

To calibrate more accurately the photoelastic resonances based on non-resonant values reported in Ref. [13, 14], we extended the model of the photoelastic coefficient presented in Ref. [22] to our strain configuration and applied it to fit the data of [13, 14] so as to extrapolate its dependence to higher energies. The result is shown by the dotted line in Fig.3, to be compared with our measurement at 290K. There is a clear difference between the two curves well below the exciton resonance, a reasonable observation as our model only contains one resonant contribution (that of the heavy hole to conduction electron excitonic transition) and no slowly varying contribution to the dielectric constant from other higher energy optical transitions. As this contribution is very small as compared to the resonant contribution considered in this work (note the logarithmic scale), it is sufficient to consider the range of energy where the two models overlap (-20 to -5 meV typically) to determine the absolute values of the photoelastic modulus at resonance, as derived in our experiment. The lines corresponding to the different measured temperatures in Fig.3 were obtained with this calibration. The maximum photoelastic constant attained at resonance is displayed as a function of the temperature in the bottom right of Fig.3. It ranges from 33 to 53000 between 290 and 30K. Even when corrected by the GaAs filling factor in the structure, it remains larger by 2 to 5 orders of magnitude as compared to the non-resonant value in GaAs (0.14) used in Ref. [6] to estimate optomechanical coupling in GaAs based nanostructures.

In conclusion, we have demonstrated huge polariton-mediated resonant enhancement of the photoelastic coupling in GaAs/AlAs MQWs (up to $10^5$). We have shown that the resonant Brillouin scattering is a powerful tool that allows one to quantitatively determine photoelastic coupling amplitudes between light and ultrasound in semiconductor nanostructures across excitonic transitions over a large temperature range between 30 and

290K. We have shown that in a GaAs/AlAs multiple quantum well the energy dependence of $p_{12}$ across the heavy-hole exciton resonance can be described within a polariton scattering picture involving three material related parameters only. A constant radiative damping rate well describes the results in the whole temperature range while the non-radiative damping rate increases from 70 µeV to 2.9 meV with growing temperature. Brillouin measurements provide relative variations of the photoelastic coupling amplitude as a function of energy and temperature and we calibrated the absolute amplitude based on piezo-optics experiments in the transparency region. Our results demonstrate the great advantage to extend nano-optomechanical experiments towards excitonic transition energies with expected several order of magnitude increases of optomechanical coupling. They give some quantitative support to the recent proposals towards a new polariton optomechanics [25, 26]. We suggest multiple quantum wells as an excellent option for high sensitivity resonators. Based on the geometry and the non resonant values given in Ref. [6], optomechanical coupling factor $g_0$ in the range of 0.1 to 100 THz could be expected depending on the experimental temperature.

**Acknowledgements**

We thank Florent Margaillan, Mathieu Bernard and Silbe Majrab for essential technical support. ANP and APV acknowledge the support of the Russian Foundation for Basic research and the Programs of RAS.